\documentclass[conference,a4paper]{IEEEtran}

\usepackage{cite}
\usepackage{graphicx,color,epsfig,rotating}
\usepackage{amsfonts,amsmath,amssymb,bbm}
\usepackage{algorithm}
\usepackage{algpseudocode}
\usepackage{amsmath}
\usepackage{cite}
\usepackage{mdwtab} 
\usepackage{placeins}
\usepackage{psfrag, graphicx}
\usepackage[latin1]{inputenc}
\usepackage{amssymb}
\usepackage{makeidx}
\usepackage{epstopdf}
\usepackage{enumitem}
\usepackage{pstricks}
\usepackage{subcaption}
\usepackage{caption}

\long\def\comment#1{}

\newfont{\bbb}{msbm10 scaled 700}

\newfont{\bb}{msbm10 scaled 1100}

\newcommand{\sv}{{\bf s}}

\newcommand{\Xm}{{\bf X}}

\newcommand{\Ac}{{\cal A}}

\newcommand{\Dc}{{\cal D}}

\newcommand{\Lc}{{\cal L}}
\newcommand{\Mc}{{\cal M}}
\newcommand{\Nc}{{\cal N}}
\newcommand{\Oc}{{\cal O}}
\newcommand{\Pc}{{\cal P}}
\newcommand{\Qc}{{\cal Q}}

\newcommand{\Sc}{{\cal S}}

\newcommand{\Uc}{{\cal U}}

\newcommand{\Vc}{{\cal V}}

\newcommand{\Zc}{{\cal Z}}

\renewcommand{\arg}{{\hbox{arg}}}

\newcommand{\eqdef}{\stackrel{\Delta}{=}}

\newcommand{\be}{\begin{equation}}
\newcommand{\ee}{\end{equation}}
\newcommand{\bea}{\begin{eqnarray}}
\newcommand{\eea}{\end{eqnarray}}

\newtheorem{defn}{Definition}
\newtheorem{example}{Example}
\newtheorem{theorem}{Theorem}
\newtheorem{lemma}{Lemma}
\newtheorem{proof}{Proof}

\begin{document}



\title{Decentralized Uncoded Storage Elastic Computing with Heterogeneous Computation Speeds}

\author{Wenbo Huang$^{1}$, Youxu Dong$^{1}$, Kai Wan$^{1}$, and Mingyue Ji$^{2}$
\thanks{The authors are with the Department of Electrical Engineering,
University of Utah, Salt Lake City, UT 84112, USA. (e-mail: nicholas.woolsey@utah.edu, rchen@ece.utah.edu and mingyue.ji@utah.edu)}
}

\author{
    \IEEEauthorblockN{ Wenbo Huang$^{1}$, Xudong You$^{1}$, Kai Wan$^{1}$, Robert Caiming Qiu$^{1}$, and Mingyue Ji$^{2}$ }
	\IEEEauthorblockA{  $^1$Huazhong University of Science and Technology, 430074  Wuhan, China\\ $^2$University of Utah, Salt Lake City, UT 84112, USA\\
		Emails: \{eric\_huang,  xudong\_you,  kai\_wan,caiming\}@hust.edu.cn,  mingyue.ji@utah.edu}

}

\maketitle

\thispagestyle{empty}
\pagestyle{empty}


\begin{abstract}
 THIS PAPER IS ELIGIBLE FOR THE STUDENT PAPER AWARD.
Elasticity plays an important role in modern cloud computing systems. Elastic computing allows virtual machines (i.e., computing nodes) to be preempted when high-priority jobs arise, and also allows new virtual machines   to participate in the computation. 
In 2018, Yang et al. introduced Coded Storage Elastic Computing (CSEC) to address the elasticity using coding technology, with lower storage and computation load requirements. However, CSEC is limited to certain types of computations (e.g., linear) due to the coded data storage based on linear coding.
Then 
   Centralized Uncoded Storage Elastic Computing (CUSEC)    with heterogeneous computation speeds was proposed,  which  directly copies parts of data into the virtual machines.  
 In all existing works in elastic computing, the storage assignment is centralized, meaning that  the number and identity of all virtual machines possible used in the whole computation process are known  during the storage assignment. 
In this paper, we consider   Decentralized Uncoded Storage Elastic Computing (DUSEC)  with heterogeneous computation speeds, where 
  any available virtual machine can join the computation which is not predicted and thus   coordination among different virtual machines'
storage assignments  is not allowed.   
  Under a decentralized storage assignment originally proposed in coded caching by  Maddah-Ali and Niesen, we propose a computing scheme with closed-form optimal computation time.
We also run experiments  over MNIST dataset with Softmax regression model through the Tencent cloud platform, and the experiment results demonstrate that the proposed DUSEC system approaches the state-of-art best storage assignment in the CUSEC system in computation time.
\end{abstract}



\section{Introduction}
\label{section: intro}

Cloud computing platforms provide elastic computation service at discount, while the computations are scheduled on the Virtual Machines (VMs) at a low-priority. It means that at each time step of the computation process (i) VMs will be preempted if a high-priority job arrives; (ii) new available VMs are allowed to join the computation at any time~\cite{meng2016mllib,abadi2016tensorflow,yang2018coded}. To efficiently tolerate the failures brought by preempted VMs, 
Yang et al~\cite{yang2018coded} introduced Coded Storage Elastic Computing (CSEC) to address the elasticity using coding technology, with lower storage and computation load requirements. 
Following the original CSEC work, various works on the extensions such as   elastic computing with heterogeneous storage or/and speed,   elastic computing against stragglers, optimization on the transition waste, were proposed in~\cite{Kiani_Hierarchical,woolsey2021cec,woolsey2021practicalcec,CEC2023zhong,dau2020optimizing}.
Despite the advantages of CSEC such as less storage overhead, it may be challenging to be applied to more involved computations (e.g.,non-linear task, deep learning) due to the coded data storage. So we may prefer to place the data in an uncoded way by just assigning the raw data to the VMs. \cite{Ji_Uncoded} proposed a framework for heterogeneous Uncoded Storage Elastic Computing (USEC) in Matrix-vector computation and~\cite{zhong2024uncoded} considered a Matrix-Matrix computation task in uncoded storage systems. 

To the best of our knowledge, in all existing works on elastic computing, the storage assignment is centralized, meaning that  the number and identity of all virtual machines possible used in the whole computation process are known  in prior during the storage assignment.
In this paper we consider   Decentralized Uncoded Storage Elastic Computing (DUSEC), 
 where the identity of the VMs participating into the computing process is not in prior known at the beginning of the whole computation  process. In other words,
 any available virtual machine can join the computation and thus   coordination among different virtual machines' storage assignments  is not allowed. 
Unlike the centralized system,  there is no limit on the number of available VMs $N$ at time step $t$ in the DUSEC system; as $N$ increases, the computation time decreases. 
For the computation task, we consider the   linearly separable function~\cite{Damon2008fast}, which is  a    function of $K$ datasets ($D_1,\ldots,D_{K}$) on a finite field $\mathbb{F}_{q}$. 
The task function can be seen as    $K^{(t)}_{c}$ linear combinations of $K$ intermediate messages. 
It was shown in~\cite{wan2021distributed} that such function could cover Matrix-matrix multiplication,  gradient descent, linear transform, etc., as special cases.  
In addition, we consider that VMs have   heterogeneous computation speeds, and aim to minimize the computation time at time $t$ defined as the largest computation time among all available VMs at time step $t$.   

For this new problem,   referred to as   DUSEC    with heterogeneous computation speeds,  
we consider the case $K^{(t)}_{c}=1$, which covers matrix-vector multiplication and gradient descent tasks, and 
our main contribution is summarized as follows:
\begin{enumerate}
  \item 
We use the decentralized storage assignment originally proposed in decentralized coded caching~\cite{maddah2015decentralized}, where each VM randomly selects a fraction of datasets to store when it  joins in the computation. By assuming the number of datasets is large enough, the  storage assignment   is symmetric (i.e., the number of datasets stored inclusively stored by a set of VMs only depends on the cardinality of this set). Considering the heterogeneous computation speeds of the VMs, We formulate the computation   assignment into a convex optimization problem to achieve the minimum computation time at each time step $t$.   We then  solve the minimum computation time in closed-form, and 
propose a new     algorithm  on assigning the computation   assignment while  achieving this minimum  computation time. Note that the algorithm complexity is $\Oc(N)$, linear with the number of available VMs at time step $t$, while the classic algorithm to solve this convex optimality has complexity   $\Oc(2^{N})$ and cannot provide a close-form solution.
  \item We perform experiments through real cloud platform  with heterogeneous computation speeds, and run over the MINST dataset with a Softmax model. We demonstrate that in terms of the total processing time, the proposed algorithms on DUSEC   approaches   the start-of-the-art CUSEC scheme in~\cite{Ji_Uncoded}, which requires the knowledge on the identity of the VMs at the beginning of the process.   
  \item We then extend the proposed DUSEC scheme by using the distributed gradient coding scheme   in~\cite{heterogeneous_optimal}, such that the resulting elastic computing scheme can also tolerate potential stragglers in the computation process.\footnote{\label{foot:elastic}The difference between elasticity and straggler is that, for elasticity at the beginning of each step time  we know the identity of the computing nodes who has joined in or left; but  we do not know which nodes will be stragglers.} 
\end{enumerate}

\paragraph*{Notation Convention}
We use $|\cdot|$ to represent the cardinality of a set or the length of a vector
and $[n] \eqdef \{1,2,\ldots,n\}$. 
A bold symbol such as $\boldsymbol{a}$ indicates a vector and $a[i]$ denotes the $i$-th element of $\boldsymbol{a}$. 
Calligraphic symbols such as $\Ac$ presents a set with numbers as its elements. Bold calligraphic symbols such as $\boldsymbol{\Ac}$ represents a collection of sets. 

\section{Problem Formulation}
\label{sec: Network Model and Problem Formulation}
A server uses VMs in a cloud to perform  linearly separable computation tasks over multiple time steps. 
At each time step $t$, 
the computation task is  a  function of $K$  datasets $D_1, \ldots, D_{K}$, which should be computed collaboratively by    $N_t$ available VMs in $\Nc_t$ with $N_t:= |\Nc_t|$. 
As in~\cite{wan2021distributed}, with the assumption that the function is linearly separable from the datasets, the computation task can be   written as $K^{{(t)}}_{c}\leq K$ linear combinations of $K$ messages,
\begin{align}
    &f^{(t)}({D_1},{D_2}, \ldots ,{D_{K}}) = g^{(t)}({f^{(t)}_1}({D_1}), \ldots ,{f^{(t)}_{K}}({D_{K}}))  \nonumber\\
    &= g^{(t)}({W^{(t)}_1}, \ldots ,{W^{(t)}_{K}}) = {\bf F}^{(t)}[{W^{(t)}_1}; \ldots ;{W^{(t)}_K}], \label{eq:computation task}
\end{align}
where the $i^{\text{th}}$ message is  ${W^{(t)}_i} = f_i^{(t)}({D_i}) $, representing  the outcome of the  component function $f^{(t)}_i(\cdot)$ applied to dataset $D_i$, and ${\bf F}^{(t)}$ represents the demand information matrix with dimension $K^{{(t)}}_{c} \times K$, known by all VMs. 
Each message $W^{(t)}_i$ contains $L$ uniformly i.i.d. symbols on some finite field $\mathbb{F}_{q}$.\footnote{\label{foot:real}The proposed scheme can also work in the field of real numbers.}

In this paper, we mainly consider the case $K^{(t)}_c = 1$, which covers matrix-vector multiplication and gradient descent tasks (by letting $f_i^{(t)}$ be a gradient of loss function). In this case, we assume without loss of generality that the computation task at time step $t$ is $W^{(t)}_1 + \cdots +W^{(t)}_K$.

\subsection{Decentralized Storage Assignment and Heterogeneous Computation speeds}
We consider decentralized system, where any VM may join the computation process unpredictably and thus coordination on storage assignment for different VMs is not allowed. 
We use the decentralized storage assignment in~\cite{Maddah_decentralized}, where each VM stores a subset of the datasets independently at random. 
Assume that each VM $n$ stores $\{D_i: i\in \Zc_n\}$, with a equal storage size $|\Zc_n| = M$.
So   each dataset is stored by each VM with probability $\frac{M}{K}$.

Since there is no coordination among VMs' storage assignment and the computation tasks at different time steps are independent, in the rest of this paper we only focus on one time step $t$; and
to avoid heavy notations, 
we do not explicitly point out the time step index $t$ in the notations. For example, we drop the superscript from $W^{(t)}_i$ and the $i^{\text{th}}$ message becomes $W_i$. We further assume that the available VMs at the considered time step is $\Nc=[N]$.

At the considered time step, according to the storage of the VMs in $[N]]$, we can divide the $K$ datasets into $2^{N}$ sets, where 
${\Ac}_{\Vc}$ represents the sets of datasets   assigned to all VMs in $\{V_j:j\in \Vc\}$, for each $\Vc \subseteq [N]$. 
By assuming that $K$ is large enough and then by the law of large numbers, we have 
\begin{align}
    \label{eq: A_r}
    |{\Ac}_{\Vc}| = \left(\frac{M}{K}\right)^{|\Vc|}\left(1-\frac{M}{K}\right)^{N - |\Vc|}K  +o(K),
\end{align}
with probability approaching one when $K$ is sufficiently large. 

Note that by the  decentralized storage assignment, there exist  some datasets not stored by any VMs in $[N]$; thus we can only compute an approximated version of the original computation task. So we change the computation task to  $\boldsymbol{y} = \sum_{i \in \bigcup_{j\in [N]}\Zc_{j}} W_{i}. $

We consider the scenario  of heterogeneous computation speeds, where each VM $V_j$ where $j\in [N]$ has the computation speed $s[j]$.
Without loss of generality, we assume that the computation speeds are in an ascending order as $s[1]\le s[2]\le \cdots \le s[N]$. Denote $\boldsymbol{s}=(s[1],\ldots,s[N])$.  

Next, we  define $\Lc_{n}$ as the set of datasets 
 which are   only assigned to some of the first $n$ VMs; thus 
\begin{equation}
\Lc_{n} = \Ac_{{1}} \cup \cdots \cup \Ac_{{n}} \cup \Ac_{\{1,2\}} \cup \cdots \cup \Ac_{\{1,2,\ldots, n\}} .
\end{equation}
By~\eqref{eq: A_r},  we have with high probability that
\begin{align}
    |{ \Lc}_{n}| = &\left(\frac{K - M}{K}\right)^{N}\left\{\left(\frac{K}{K - M}\right)^{n} - 1\right\}K +o(K).  
    \label{eq:Ln length}
\end{align}
To simply the notation, define $\alpha := \frac{K}{K - M}$ and $\beta:= \left(\frac{K-M}{K}\right)^{N} $. Then we have
\begin{align}
    |{\Ac}_{\Vc}|/K = \beta \left(\alpha - 1\right)^{|\Vc|}, \ |{\Lc}_{n}| /K= \beta \left({\alpha}^{n} - 1\right).
    \label{eq:normalized Av and Lv}
\end{align}


\subsection{USEC under Decentralized Assignment}

Each  $V_n$ where $n \in [N]$  computes $\sum\limits_{\Vc\subseteq [N]: n \in \Vc} \sum\limits_{i \in \Sc_{n,\Vc}} W_{i}$, where $\Sc_{n,\Vc} \subseteq \Ac_{\Vc}$ denotes the set of datasets in $\Ac_{\Vc}$ which should be computed by VM $V_n$. 
So the computation assignment for all VMs can be exactly determined by  the collection,  $\boldsymbol{M} = \left\{\Sc_{n,\Vc}, \forall n \in [N], \forall \Vc \subseteq  [N] \right\}.$

Then  the computed results from VMs are transmitted to the server   to recover $\boldsymbol{y}$.  


{\bf Computation Load.} 
We define $\mu[n]$, the computation load by each VM $V_n$, as the number of computed messages normalized by $K$; thus we have 
\begin{align}
    \mu[n]=\sum_{\Vc\subseteq [N]: n\in \Vc} \frac{|\Sc_{n,\Vc}|}{K}=\sum_{\Vc\subseteq [N]: n\in \Vc} \mu[n,\Vc],
    \label{eq: compload_vector}
\end{align}
where we define  $\mu[n,\Vc]=|\Sc_{n,\Vc}|/K$.
Then the computation load vector for the VMs in $[N]$ is,  $\boldsymbol{\mu} = \left(\mu[1], \cdots, \mu[n]\right)$. 
Note that since the computation load is normalized by $K$, thus we can neglect the deviation terms and assume that 

{\bf Computation Time.}
The computation time  of each VM $V_n$ where $n\in [N]$ is $\frac{\mu[n]}{s[n]}$. The overall computation time 
at the considered time step (or simply called computation time) is defined as largest computation time among all VMs in $[N]$, 
\begin{align}
\label{eq: comptime} 
c(\boldsymbol{M})  \eqdef \max_{n \in [N]} \frac{\mu[n]}{s[n]} = \max_{n \in [N]} \frac{\sum_{\Vc \subseteq [N]: n\in \Vc }\mu[n,\Vc]}{s[n]}. 
\end{align}

Without considering stragglers, 
each message  only needs to be computed once. For a fixed storage assignment $(\Zc_1,\ldots,$ $\Zc_N)$, we can formulate the following optimization problem  for the DUSEC system with heterogeneous computation speeds:
\begin{subequations} \label{eq: uncoded opt}
\begin{align}
\underset{\boldsymbol{M}}{\text{minimize}} & \quad c\left(\boldsymbol{M}\right) \label{eq: uncoded 1} \\
\text{subject to:}  &\bigcup_{n \in \Vc} \mathcal{S}_{n,\Vc} =  \Ac_{\Vc}, \ \forall \Vc \subseteq \Nc \label{eq: unocded constraint 1}. 
\end{align}
\end{subequations} 
The optimization problem in~\eqref{eq: uncoded opt} is  equivalent to the following convex optimization problem:
\begin{subequations} \label{eq: uncoded opt 2}
\begin{align}
 \underset{\{\mu[n,\Vc]: n\in [N], \Vc\subseteq [N]\} }{\text{minimize}} &  c\left(\boldsymbol{M}\right) = \max_{n \in  [N]}  \frac{\sum_{\Vc \subseteq [N]: n\in \Vc }\mu[n,\Vc]}{s[n]} \label{eq: uncoded 2} \\
\text{subject to:} &  \quad \sum_{n \in  \Vc} \mu[n,\Vc] = \beta \left(\alpha - 1\right)^{|\Vc|} , \ \forall \Vc \subseteq [N], \label{eq: usec constraint mu 2}\\ 
&   \mu[n,\Vc] \geq 0, \forall n \in [N], \Vc \subseteq [N].
\end{align}
\end{subequations}

\section{Main Results}
Instead of directly solving the convex optimization problem in~\eqref{eq: uncoded opt 2} by the method of Lagrange multipliers or other standard methods, we solve~\eqref{eq: uncoded opt 2} by first proving a cut-set converse bound on the computation time and then  
proposing an algorithm which matches the converse bound. Thus we can 
 provide the optimal computation time in closed-form  for the optimization problem in~\eqref{eq: uncoded opt 2}, which is shown in the following theorem. 
\begin{theorem}[Optimal Computation Time]
\label{thm: explicit decentralized}
    Under the  DUSEC system with heterogeneous computation speeds, 
     the optimal computation time is 
    \begin{subequations}
    \begin{align}
       & c^{\star}= \min\limits_{\boldsymbol{M}} c(\boldsymbol{M}) = \underset{{{n^{*}}}}{{\max}} \frac{|\Lc_{n^{*}}|/K}{\sum_{i=1}^{n^*}s[i]}, \\
       & \text{where} \quad  \frac{|\Lc_{n^{*}}|/K}{\sum_{i=1}^{n^*}s[i]} \ge 
        \frac{|\Lc_{n} \setminus \Lc_{n^{*}}|/K}{\sum_{i=n^*+1}^{n}s[i]}, \  \forall n \in [n^*+1 : N ],  \tag{10b} \label{cond:1}\\
       & \text{and}   \quad \frac{|\Lc_{n^{*}}|/K}{\sum_{i=1}^{n^*}s[i]} \ge 
        \frac{|\Lc_{n}|/K}{\sum_{i=1}^{n}s[i]}, \forall n \in [n^*-1].        \tag{10c} \label{cond:2}
    \end{align}
    \end{subequations}
\end{theorem}
We first show that there must exist some $n^{*} \in [N]$ satisfying the constraints in~\eqref{cond:1} and~\eqref{cond:2}.  
\begin{lemma}
    \label{lem:existence of $c^*$}
There exists a $n^*$ that satisfies the Condition~\eqref{cond:2} and Condition~\eqref{cond:1} in Theorem~\ref{thm: explicit decentralized}.
\end{lemma}
The proof of Lemma~\ref{lem:existence of $c^*$} is given in Appendix~\ref{appendix: exsitence}. Then the proof of the converse bound   in Theorem~\ref{thm: explicit decentralized} is given in Appendix~\ref{sec:converse proof}.

\subsection{Achievability Proof of Theorem~\ref{thm: explicit decentralized}}
\label{sub:achievable}
We first provide an example to illustrate the main idea of the proposed scheme which achieves the optimal computation time in Theorem~\ref{thm: explicit decentralized}. Recall that our main objective is to fix the set of messages computed by each worker $n\in [N]$, i.e.,  the set  $\Sc_{n} \eqdef \bigcup_{\Vc \subset [N]: n \in \Vc} \Sc_{n,\Vc}$. 

\begin{example}
We consider a system with parameters $(K = 16000, N = 4, \alpha = 2, \sv = [1,2,5,5]).$ There are $K = 16000$ datasets, and each VM can store $M = 8000$ datasets. By the law of the large number, we have $|\Lc_{1}| / K =  0.0625, |\Lc_{2}| / K = 0.1875 , |\Lc_{3}| /K = 0.4375, |\Lc_{4}| / K = 0.9375.$ Note that from Theorem~\ref{thm: explicit decentralized}, we can get the $n^{*} = 4$ and $c(\boldsymbol{M}) = 0.0721$. We present the computation load assignment to achieve  $c(\boldsymbol{M}) = 0.0721$ as follows.
\begin{figure}
\centering
\includegraphics[width=0.8\columnwidth]{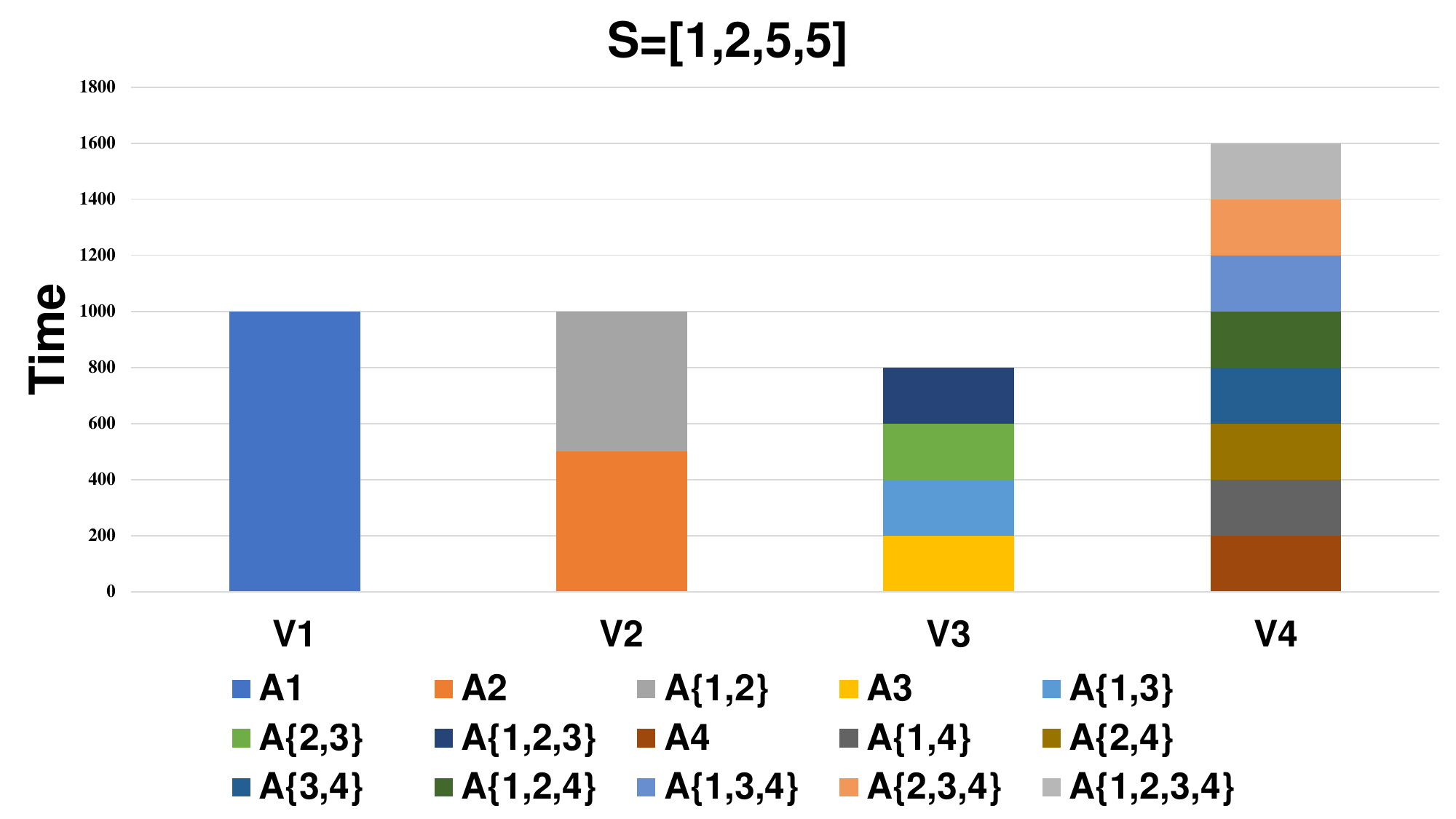}
\caption{Illustration of  DUSEC when $\sv = [1,2,5,5]$ before rearrangement, $c(\boldsymbol{M}) = 0.1$}
\label{fig: Uncoded_Elastic_Computing_[1,2,5,5]_1} 
\end{figure}

There are $4$ iterations in the algorithm to get  the  computation  load assignment $\{\Sc_1, \Sc_2, \Sc_3, \Sc_4\}$ for $N = 4$ VMs. In the $n$-th iteration, we range the computation load for the first $n$ VMs with datasets in $\Lc_n$.  

In the $1$st iteration, $V_1$ computes $\Lc_{1} = \Ac_{1}$ in $ t_1 = 0.0625.$ We update $t_1 = 0.0625,$ and $\Sc_1 = \Ac_1.$

In the $2$nd iteration, we assign $\Lc_2 \setminus  \Lc_1 = \Ac_{2}\cup\Ac_{\{1,2\}}$ to $V_2$, we have $t_2 = \frac{|\Lc_2 \setminus  \Lc_1| / K}{s[2]} = 0.0625$ which is equal to $t_1.$ We update $\Sc$ into $\Sc_{1} = \Ac_{1}, \Sc_{2} = \Ac_{2}\cup\Ac_{\{1,2\}},$ and $t_1 = t_2 = 0.0625.$ 

In the $3$rd iteration, we assign $\Lc_3 \setminus  \Lc_2 = \Ac_{3}\cup\Ac_{\{1,3\}} \cup\Ac_{\{2,3\}} \cup\Ac_{\{1,2,3\}}$ to $V_3$, we have $t_3 = \frac{|\Lc_3 \setminus  \Lc_2| / K}{s[3]} = 0.05$ while $t_3 < t_2 = t_1$ satisfies Condition~\ref{cond:1} in Theorem~\ref{thm: explicit decentralized}. We update $\Sc$ into $\Sc_{1} = \Ac_{1}, \Sc_{2} = \Ac_{2}\cup\Ac_{\{1,2\}}, \Sc_{3} = \Ac_{3}\cup\Ac_{\{1,3\}}\cup\Ac_{2,3}\cup\Ac_{\{1,2,3\}},$ and $t_3 = 0.05 < t_1 = t_2 = 0.0625.$ 

In the $4$th iteration, we assign $\Lc_{4} \setminus \Lc_{3}$ to $V_4$ firstly, then $c(\boldsymbol{M}) = t_4 = \frac{|\Lc_{4} \setminus \Lc_{3}| / K}{s[4]} = 0.1 > c^{\star} > t_3$, which contradicts with Condition~\ref{cond:1} in Theorem~\ref{thm: explicit decentralized}. 

To reduce $c(\boldsymbol{M}) = t_4,$ the datasets which are stored in both $V_{3}$ and $V_{4}$ should be computed by both of them. With $t_3= 0.05, t_4 =0.1,$ we compute $ t_{\{3,4\}} = \frac{|\Lc_{4} \setminus \Lc_{2}| / K}{s[3]+s[4]} = 0.075,$ and rearrange the computation load of $V_3.$ $V_3$ need to compute datasets from  $\Ac_{\{3,4\}} \cup \Ac_{\{1,3,4\}} \cup \Ac_{\{2,3,4\}} \cup \Ac_{\{1,2,3,4\}} $ with size $\delta =  \left(t_{\{3,4\}} - t_{3}\right)s[3] = 0.125$. 

The main non-trivial technology is that the original computation load of $V_3$ are  $\Sc_3 = \Ac_{3} \cup \Ac_{\{1,3\}} \cup \Ac_{\{2,3\}} \cup \Ac_{\{1,2,3\}},$ while
$\{\Ac_{3}, \Ac_{\{1,3\}}, \Ac_{\{2,3\}}, \Ac_{\{1,2,3\}}\}$ are one-to-one corresponding to $\{\Ac_{\{3,4\}},\Ac_{\{1,3,4\}}, \Ac_{\{2,3,4\}}, \Ac_{\{1,2,3,4\}}\}.$ Similarly, the original computation load of $V_4,$  $\Lc_4 \setminus \Lc_3$ are one-to-one corresponding to $\{\Ac_{\{3,4\}},\Ac_{\{1,3,4\}}, \Ac_{\{2,3,4\}}, \Ac_{\{1,2,3,4\}}\}.$ From the size of rearranged computation load $\delta = 0.125$ and $\mu[3] = 0.25,$ $\mu[4] = 0.5,$
we can update $\Sc_3$ into 
\begin{subequations}
\begin{align*}
    &\Sc_3 = \Ac_{3} \cup \Ac_{\{1,3\}} \cup \Ac_{\{2,3\}} \cup \Ac_{\{1,2,3\}} \\
    &\cup 0.5\Ac_{\{3,4\}} \cup 0.5\Ac_{\{1,3,4\}} \cup 0.5\Ac_{\{2,3,4\}} \cup 0.5\Ac_{\{1,2,3,4\}},
\end{align*}
\text{as well as,}
\begin{align*}
    &\Sc_4 = \Ac_{4} \cup \Ac_{\{1,4\}} \cup \Ac_{\{2,4\}} \cup \Ac_{\{1,2,4\}} \\
    &\cup 0.5\Ac_{\{3,4\}} \cup 0.5\Ac_{\{1,3,4\}} \cup 0.5\Ac_{\{2,3,4\}}  \cup 0.5\Ac_{\{1,2,3,4\}}.
\end{align*}
\end{subequations}
After the computation load rearrangement, $c(\boldsymbol{M}) = t_{3} = t_4 = 0.075 > t_1$ while the maximum time is not achieved by $V_1$ which contradicts with Condition~\ref{cond:1} in Theorem~\ref{thm: explicit decentralized}
see in Fig.~\ref{fig: Uncoded_Elastic_Computing_[1,2,5,5]_2}.

We further rearrange the computation load between $\{V_1,V_2\}$ and $\{V_3,V_4\}$.
\begin{figure}
\centering
\includegraphics[width=0.8\columnwidth]{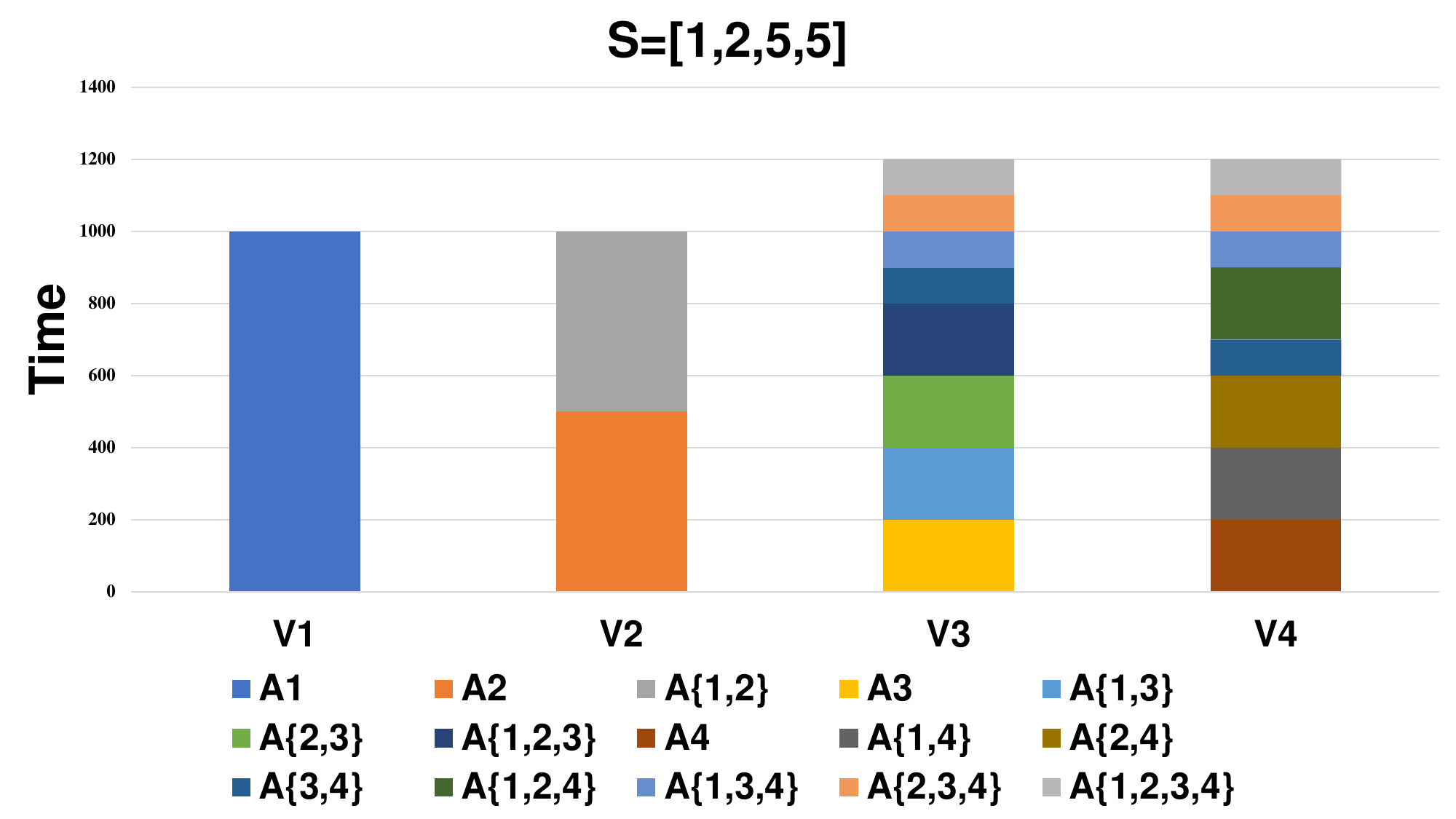}
\caption{Illustration of  DUSEC when $\sv = [1,2,5,5]$, $c(\boldsymbol{M}) = 0.075$ }
\label{fig: Uncoded_Elastic_Computing_[1,2,5,5]_2} 
\end{figure}
With $t_1 = t_2 = t_{\{1,2\}}= 0.0625, t_3 = t_4 = 0.075, t_{\{1,2,3,4\}} = 0.0721$, the common part of datasets between $\{V_1, V_2\}$ and $\{V_3, V_2\}$ are
\begin{subequations}
\begin{align*}
     \{\Ac_{\{1,3\}} \cup \Ac_{\{1,4\}} \cup \Ac_{\{1,3,4\}},\\
     \Ac_{\{2,3\}} \cup \Ac_{\{2,4\}} \cup \Ac_{\{2,3,4\}},\\
     \Ac_{\{1,2,3\}} \cup \Ac_{\{1,2,4\}}\cup \Ac_{\{1,2,3,4\}}\},
\end{align*} 
\end{subequations}
which can be one-to-one corresponding to $\{\Ac_{1}, \Ac_{2}, \Ac_{\{1,2\}}\}$ with size $\delta  = 0.028875.$ In another division 
\begin{subequations}
\begin{align*}
    \{\Ac_{\{1,3\}} \cup \Ac_{\{2,3\}} \cup \Ac_{\{1,2,3\}}, \\
    \Ac_{\{1,4\}} \cup \Ac_{\{2,4\}} \cup \Ac_{\{1,2,4\}}, \\
    \Ac_{\{1,3,4\}} \cup \Ac_{\{2,3,4\}} \cup \Ac_{\{1,2,3,4\}} \}
\end{align*}
\end{subequations} are one-to-one corresponding to $\{\Ac_{3}, \Ac_{4}, \Ac_{\{3,4\}}\}$ with size $\delta = 0.028875$. 

Depending on the correlation of common storage part and original computation load, the rearranged computation load can be designed as
$\Ac'_{n,\Vc\cup\Qc}, n \in [2], \forall \Vc \subset [2], \Qc \subset [3:4],$ where $|\Ac'_{n,\Vc\cup\Qc}|$ is corresponding to $\mu[n,\Vc]$ in the original computation load of $\{V_1,V_2\},$ and $|\Ac_{\Qc}|$ in $|\Lc_2|.$ And the summation of $|\Ac'_{n,\Vc\cup\Qc}|,n \in [2], \forall \Vc \subset [2], \Qc \subset [3:4],$ is equal to $\delta.$ The computation load arrangement can be designed as $\Ac'_{n,\Vc\cup\Qc} = \frac{\delta |\Ac_{\Qc}|\mu[n,\Vc]}{|L_2||L_2||\Ac_{\Vc \cup \Qc}|}\Ac_{\Vc \cup \Qc}, n \in [2],$  and $\Ac'_{n,\Vc\cup\Qc} =\frac{\delta|\Ac_{\Vc}| \mu[n,\Qc]}{|L_2||L_2||\Ac_{\Qc \cup \Vc}|}\Ac_{\Qc \cup \Vc}, n \in [3:4].$

The computation load rearranged to $\Sc_n, n \in [2]$ are
\begin{subequations}
\begin{align*}
  \Sc_{n,\Vc \cup \Qc} &= \frac{\delta |\Ac_{\Qc}|\mu[n,\Vc]}{|L_2||L_2||\Ac_{\Vc \cup \Qc}|}\Ac_{\Vc \cup \Qc}, \\
  &\forall \Vc \subset [2], \Qc \subset [3:4],  
\end{align*}
\end{subequations}
and the computation load rearranged to $\Sc[n], n \in [3:4]$ are
\begin{subequations}
\begin{align*}
    \Sc_{n,\Qc\bigcup \Vc} &= \Sc[n,\Qc \cup \Vc] \setminus \frac{\delta|\Ac_{\Vc}| \mu[n,\Qc]}{|L_2||L_2||\Ac_{\Qc \cup \Vc}|}\Ac_{\Qc \cup \Vc}, \\
    &\forall \Qc \subset[3:4],  \Vc \subset [2].
\end{align*}
\end{subequations}

The optimal computation load assignment of  $\Sc_{1}, \Sc_{2}, \Sc_{3}, \Sc_{4}$ is presented in the Table~\ref{tab:exchanged_[1,2,5,5]}.

\begin{figure}
\centering
\includegraphics[width=0.8\columnwidth]{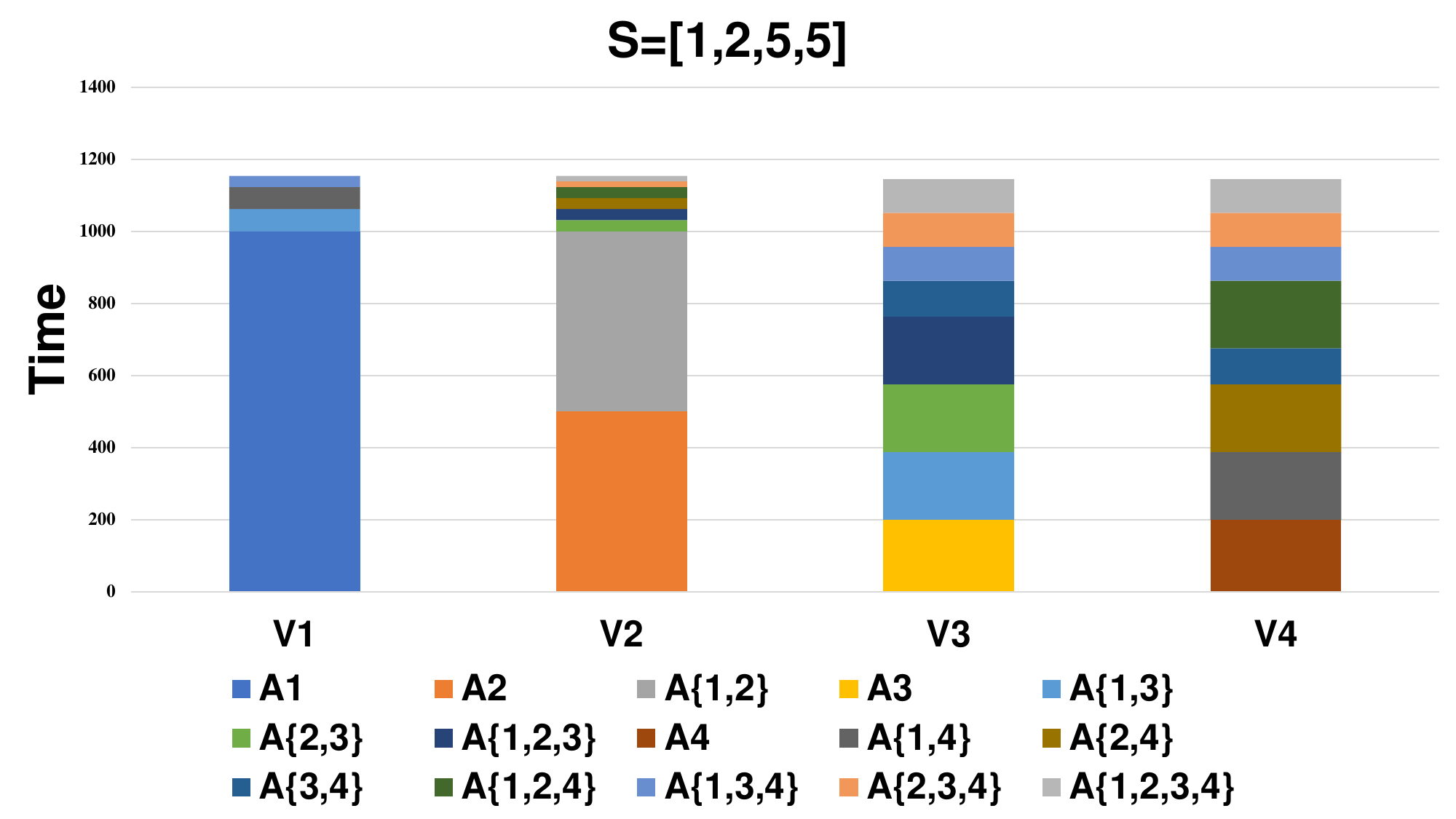}
\caption{Illustration of  DUSEC when $\sv = [1,2,5,5]$ after rearrangement, $c^{\star} = c(\boldsymbol{M}) = 0.0721$}
\label{fig: Uncoded_Elastic_Computing_[1,2,5,5]_3} 
\end{figure}

\begin{table}
  \centering
  \caption{The computation load assignment for $s=[1,2,5,5]$ system}
  \label{tab:exchanged_[1,2,5,5]}
  \begin{tabular}{c|c|c|c|c}
    \hline
    \textbf{} & \textbf{$\Sc_1$} & \textbf{$\Sc_2$} & \textbf{$\Sc_3$} & \textbf{$\Sc_4$} \\
    \hline
    $\Ac_{1}$ & 1 & 0 & 0 & 0 \\
    \hline
    $\Ac_{2}$ & 0 & 1 & 0 & 0 \\
    \hline
    $\Ac_{3}$ & 0 & 0 & 1 & 0 \\
    \hline
    $\Ac_{4}$ & 0 & 0 & 0 & 1 \\
    \hline
    $\Ac_{\{1,2\}}$ & 0 & 1 & 0 & 0 \\
    \hline
    $\Ac_{\{1,3\}}$ & 0.0616 & 0 & 0.9384 & 0 \\
    \hline
    $\Ac_{\{1,4\}}$ & 0.0616 & 0 & 0 & 0.9384 \\
    \hline
    $\Ac_{\{2,3\}}$ & 0 & 0.0616 & 0.9384 & 0 \\
    \hline
    $\Ac_{\{2,4\}}$ & 0 & 0.0616 & 0 & 0.9384 \\
    \hline
    $\Ac_{\{3,4\}}$ & 0 & 0 & 0.5 & 0.5 \\
    \hline
    $\Ac_{\{1,2,3\}}$ & 0 & 0.0616 & 0.9384 & 0 \\
    \hline
    $\Ac_{\{1,2,4\}}$ & 0 & 0.0616 & 0 & 0.9384 \\
    \hline
    $\Ac_{\{1,3,4\}}$ & 0.0308 & 0 & 0.4846 & 0.4846 \\
    \hline
    $\Ac_{\{2,3,4\}}$ & 0 & 0.0308 & 0.4846 & 0.4846 \\
    \hline
    $\Ac_{\{1,2,3,4\}}$ & 0 & 0.0308 & 0.4846 & 0.4846 \\
    \hline
  \end{tabular}
  \label{tab:example}
\end{table}
    \hfill $\square$ 
\end{example}

\paragraph{Algorithm~\ref{algorithm:decentralized 1}}
The algorithm of the computation load assignment consists of  $N$ iterations  to calculate $\Sc_{n,\Vc}, \forall n \in [N], \forall \Vc \subset [N]$ and $\Sc_{n} \eqdef \bigcup_{\Vc \subset [N]: n \in \Vc} \Sc_{n,\Vc}.$

\begin{algorithm} 
  \caption{Computation Load Assignment for Heterogeneous Computation Speed}
  \label{algorithm:decentralized 1}
  \begin{algorithmic}[1]
  \item[ {\bf Input}: $\sv$, $N$, $\alpha$ ] 
   \hspace*{4cm} 
  \For {$n \in [N]$}
  \State \quad $t_n \leftarrow \frac{|\Lc_{n} \setminus \Lc_{n-1}|/K}{s[n]} \quad $ set the computation time $t_n$.
  \State {\bf if $t_n \le t_{n-1}$} 
  \State \quad $\Sc_n \leftarrow \cup_{\Vc \subset [n]: n \in \Vc} \Ac_{n,\Vc} \quad$ set the computation load of $V_n$. 
  \State {\bf else $t_n > t_{n-1}$} 
  \State \quad $n' \leftarrow n \quad$ preliminary setting of $n'$
  \State \quad $V_{n'-1}, V_{n'-2}, \ldots, V_{x}$ find the VMs that $t_{n'-1} = \ldots = t_{x}.$
  \State \quad  $\star$  update $\{\Sc_n, \ldots, \Sc_x\}$ through Algorithm~\ref{algorithm:lemma1 1}.
  \State \quad {\bf if $ \frac{|\Lc_{n} \setminus \Lc_{x-1}|/K}{\sum^{n}_{i=x} s[i]}  \le t_{x-1} $}
  \State \quad \quad  $n \leftarrow n+1$ 
  \State \quad {\bf else $\frac{|\Lc_{n} \setminus \Lc_{x-1}|/K}{ \sum^{n}_{i=x} s[i]} > t_{x-1} $}
  \State \quad \quad  $n' \leftarrow x $ \quad {\bf reset $n'$, and return to $\star$.} 
  \State \quad {\bf end if}
  \State  {\bf end if}
  \EndFor
  \item[ {\bf Output}: $\Sc_{n},$ $ n \in {[N]}$ ]
  \end{algorithmic}
\end{algorithm}


{\bf With no rearrangement (Line 8, 9)}
For each iteration $n \in N$, in Line $7$ we take $\Lc_{n+1} \setminus \Lc_{n}$ and $V_{n+1}$ into consideration, and we first let $V_{n+1}$ compute the $\Lc_{n+1} \setminus \Lc_{n}$ by itself, when $t_{n+1} \le t_{n},$ we update $\Sc_i= \Sc_i, \forall i \in [n], \Sc_{n+1} = \Lc_{n+1} \setminus \Lc_{n},$ and $t_{n+1} = \frac{|\Lc_{n+1} \setminus \Lc_{n}|/K}{s[n+1]}.$

{\bf With rearrangement (Line 11 $\sim$ 19)}
When $t_{n+1} > t_{n},$ we try to update $t_{n+1} = t_{n}.$
Firstly, we identify the VMs $\left\{V_{n-1}, V_{n-2}, \ldots, V_{x} \right\}$ that have the same computation time as $V_{n-1}.$ Next, we update $\{\Sc_{n}, \Sc_{n-1}, \ldots, \Sc_x\}$ using Algorithm~\ref{algorithm:lemma1 1}. If $x == 1,$ we stop the rearrangement phase, if $x \neq 1,$ we repeat the process until $  t_{x-1} \ge \frac{|\Lc_{n} \setminus \Lc_{x-1}|/K}{\sum^{n}_{i=x} s[i]}$.
  
This algorithm iteratively updates the computation load assignment by considering the difference in load between consecutive time steps. It adjusts $t_{n+1}$ based on the comparison with $t_{n}$, and if necessary, applies the load balancing scheme in Algorithm~\ref{algorithm:lemma1 1} to equalize the computation times. After $N$ iterations, the algorithm converges to the final load assignment that minimizes the overall computation time $c^{\star}$ with the algorithm complexity $\Oc(N)$.

\paragraph{Algorithm~\ref{algorithm:lemma1 1}}
Algorithm~\ref{algorithm:lemma1 1} provides the detailed rearrangement for the  load balancing scheme corresponding to {\bf Line 13} in Algorithm~\ref{algorithm:decentralized 1}. In the situation $V_n, n\in [d+1:d+l]$ can compute the datasets in $\Lc_{d+l} \setminus \Lc_{d}$ in an average time $t_l = \frac{|\Lc_{d+l} \setminus \Lc_{d+1}|/K}{\sum_{n=d+1}^{d+l} s[n]}$, $V_n, n \in [d+l+1: d+l+m]$ can compute $\Lc_{d+l+m} \setminus \Lc_{d+l+1}$ in an average time $t_m = \frac{|\Lc_{d+l+m} \setminus \Lc_{d+l+1}|/K}{\sum_{n=d+l+1}^{d+l+m} s[n]},$ where $t_m > t_l.$


\begin{algorithm} 
  \caption{ Computation Load Rearrangement }
  \label{algorithm:lemma1 1}
  \begin{algorithmic}[1]
  \item[ {\bf Input}: $\boldsymbol{s}$, $d$, $l$, $m$, $\Sc_{n}$ $,n \in {[d+1 : d+l+m]} $, $t_m > t_l$, $\alpha$ ] 
   \hspace*{4cm} 
  \State  $\delta \leftarrow (\frac{|L_{l+m}|/K}{\sum_{i=1}^{l+m}s[i]} - t_{l})\sum_{i=1}^{l}s[i]$ calculate the size of rearranged computation load.
  \For {$n \in [d+1:d+l]$}
  \State  $\Sc_{n,\Vc \cup \Qc} \leftarrow \frac{ \mu[n,\Vc]|\Ac_{\Qc}|\delta}{|L_l||L_m||\Ac_{\Vc \cup \Qc}|}\Ac_{\Vc \cup \Qc} $, $ \forall \Vc \subset [l], \Qc \subset [m]$
  \EndFor
  \For {$n \in [d+l+1:d+l+m]$}
   \State  $\Sc_{n,\Vc \bigcup \Qc} \leftarrow \Sc_{n,\Vc \cup \Qc} \setminus \frac{\mu[n,\Qc]|\Ac_{\Vc}| \delta}{|L_l||L_m||\Ac_{\Qc \cup \Vc}|}\Ac_{\Vc \cup \Qc} $, $\forall \Qc \subset[m],  \Vc \subset [l]$
  \EndFor
  \item[ {\bf Output}:  $\Sc_{n},$ $ n \in {[d+1:d+l+m]}$ ]
  \end{algorithmic}
\end{algorithm}

Firstly, we calculate the average computation time $t_{l+m} = \frac{|{\Lc}_{d+l+m} \setminus \Lc_{d+1}|/K}{\sum\limits_{n \in [d+1:d+l+m]} s[n]}$ and then we calculate the size of the exchanged computation load $\delta = (t_{l+m} - t_{l})\sum\limits_{n \in [d+1:d+l]} s[n].$ 

The key technology is to divide the rearranged load of $\{V_{n}, n \in [d+1:d+l]\}$ and $\{V_{n}, n \in [d+l+1:d+l+m]\}$ into two kinds of divisions:
\begin{subequations}
    \begin{align*}
    \Dc &= \{\Ac'_{ \Vc \cup \Qc}, \forall \Vc \subset [d+1:d+l], \forall \Qc \subset [d+l+1:d+l+m]\} \\
    &= \{\Ac''_{ \Vc},  \forall \Vc \subset [d+1:d+l]\} \\  
    &= \{\Ac''_{ \Qc}, \forall \Qc \subset [d+l+1:d+l+m]\},  
\end{align*}
\end{subequations}
where $\Ac'_{\Vc \cup \Qc}$ denotes the rearranged part of $\Ac_{\Uc \cup \Vc \cup \Qc},$ $\Uc =  \emptyset \bigcup (\cup_{\forall \Pc \subset [d]} \Pc ),$ and  $\Ac''_{\Vc} $ is denoted as $ \{\Ac'_{\Vc \cup \Qc}, \forall \Qc \subset [d+l+1:d+l+m]\},$ $\Ac''_{\Qc} $ is denoted as $ \{\Ac'_{\Vc \cup \Qc}, \forall \Vc \subset [d+1:d+l]\}.$ 

We find that $\Ac''_{\Vc}$ in the rearranged part is one-to-one corresponding to $\Ac_{\Vc}$ in $\Lc_{l}$, as well as $\Ac''_{\Qc}$ in the rearranged part is corresponding to $\Ac_{\Qc}$ in $\Lc_{m}$,
so we set $|\Ac''_{\Vc}| = \frac{|\Ac_{\Vc}|}{|\Lc_{l}|}\delta,$ and $|\Ac''_{\Qc}| = \frac{|\Ac_{\Qc}|}{|\Lc_{m}|}\delta.$
As a result, $|\Ac'_{\Vc \cup \Qc}| = \frac{|\Ac_{\Vc}||\Ac_{\Qc}|}{|\Lc_{l}||\Lc_{m}|}\delta.$

We design  $\Ac'_{n, \Vc \cup \Qc}$ in the $\Ac'_{\Vc \cup \Qc}$ corresponding to $\Sc_{n,\Vc}, \forall n \in [d+1:d+l]$ in $\Ac_{\Vc}$  and  $\Ac'_{n, \Qc \cup \Vc}$ in the $\Ac'_{\Vc \cup \Qc}$ corresponding to $\Sc_{n,\Qc}, \forall n \in [d+l+1:d+l+m]$ in $\Ac_{\Qc}$. To be specifically, 
\begin{align*}
  |\Ac'_{n, \Vc \cup \Qc}| &= \frac{|\Sc_{n,\Vc}||\Ac_{\Vc}||\Ac_{\Qc}|}{|\Ac_{\Vc}||\Lc_{l}||\Lc_{m}|}\delta \\  &= \frac{\mu[n,\Vc]|\Ac_{\Qc}|}{|\Lc_{l}||\Lc_{m}|}\delta, \forall n \in [d+1:d+l],  
\end{align*}
and 
\begin{align*}
  |\Ac'_{n, \Vc \cup \Qc}| &= \frac{|\Sc_{n,\Qc}||\Ac_{\Vc}||\Ac_{\Qc}|}{|\Ac_{\Qc}||\Lc_{l}||\Lc_{m}|}\delta \\  &= \frac{\mu[n,\Qc]|\Ac_{\Vc}|}{|\Lc_{l}||\Lc_{m}|}\delta, \forall n \in [d+l+1:d+l+m].  
\end{align*}
Finally, we update $\Sc_{n}, \forall n \in [d+1:d+l]$ in {\bf Line  8} in Algorithm~\ref{algorithm:decentralized 1} as 
\begin{align*}
    \Sc_{n,\Vc \cup \Qc} &\leftarrow \frac{ \mu[n,\Vc]|\Ac_{\Qc}|\delta}{|L_l||L_m||\Ac_{\Vc \cup \Qc}|}\Ac_{\Vc \cup \Qc} , \\ &\forall \Vc \subset [l], \Qc \subset [m], \forall n \in [d+1:d+l],
\end{align*}
and 
\begin{align*}
    \Sc_{n,\Vc \bigcup \Qc} &\leftarrow \Sc_{n,\Vc \cup \Qc} \setminus \frac{\mu[n,\Qc]|\Ac_{\Vc}| \delta}{|L_l||L_m||\Ac_{\Vc \cup \Qc}|}\Ac_{\Vc \cup \Qc}, \\ &\forall \Qc \subset[m],  \Vc \subset [l], \forall n \in [d+l+1:d+l+m].
\end{align*}

\section{Evaluations on Tencent Cloud}


We evaluate the proposed algorithm using Softmax Regression 
on Tencent cloud platform. The goal is to compare the performance difference in terms of accuracy and computation time 
under DUSEC and CUSEC system.

{\it Softmax Regression}: 
Softmax regression is a model designed for solving multi-class classification problems. In Softmax regression, the model starts with a linear transformation of the input feature vector $x$ using the weight matrix $W$ and bias vector $b$ to compute scores for each class, and we train the model by updating $W$ and $b$ by $W:=W-\alpha \nabla_W J(W,b)$, where $J(\cdot)$ is the loss function. 

The network has one \verb"S5.2XLARGE16" master machine with $8$ vCPUs and $16$ GiB of memory. The worker VMs consist of $2$ \verb"S5.LARGE8" instances, each with $4$ vCPUs and $8$ GiB of memory, and $2$ \verb"S5.2XLARGE16" instances, each with $8$ vCPUs and $16$ GiB of memory. Observed that all VMs have very different computation speed, we specified the number of cores that each worker node participates in computation to simulate different speeds, normalized as $\{s[1]=1, s[2]=2, s[3] = 5, s[4] = 5\}$.

\begin{figure}
\centering
\includegraphics[width=0.75\columnwidth]{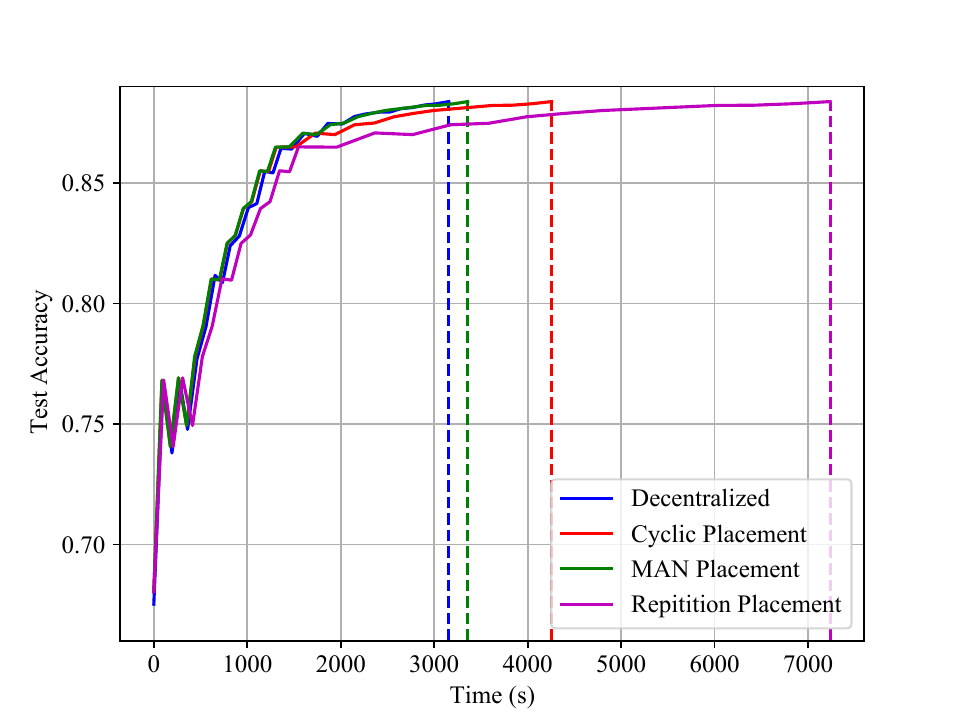}
\caption{ Results using  DUSEC and CUSEC designs on Tencent cloud platform. The y-axis represents the normalized mean square error between the true dominant eigenvector and the estimated eigenvector. 
}
\label{fig: Power}
\end{figure}

In our experiment, we conduct a comparison between  DUSEC and  CUSEC considering the presence of preemption. For the CUSEC system, we opt for repetition, cyclic,  MAN~\cite{Maddah-Ali_centralized} assignment. The results demonstrate that the proposed DUSEC system approaches the state-of-art assignment in the CUSEC system in computation time (refer to Fig.~\ref{fig: Power}). The DUSEC performs even better because of the less computation load under the decentralized assignment. 

\section{Extension to Straggler Mitigation}
\label{sec:extension}
In this section, we encode the transmission by each VM of the proposed scheme in Theorem~\ref{thm: explicit decentralized} in order to mitigate up to $s$ unpredictable stragglers in the computation process. 

Note that without elasticity, coded distributed  computing against stragglers was well studied in the literature~\cite{tandon2017gradient,ye2018communication,adaptiveGC2020,heterogeneous_optimal} to compute the linearly separable function described in Section~\ref{sec: Network Model and Problem Formulation} with $K_c=1$. In the proposed schemes~\cite{tandon2017gradient,ye2018communication,adaptiveGC2020,heterogeneous_optimal}, various coding techniques  were used to encode the messages  computed by each VM $n$ (i.e, the messages $W_j$ where $j\in \Zc_n$), who then sends the coded messages to the server. As a result, after receiving the transmissions by a fix number of VMs, the server can recover the computation task, while the communication cost is reduced compared to the uncoded transmission.  
In particular, a unified coding scheme was proposed in~\cite{heterogeneous_optimal} which works for any computation assignment $\boldsymbol{M}$ if each message is computed by at least $s+1$ VMs, in order to tolerate $s$ stragglers.

Next we provide an example to illustrate how to combine the proposed elastic scheme with the scheme in~\cite{heterogeneous_optimal}, in order to tolerate $s$ straggler, while the general description could be found in Appendix~\ref{sec:general straggler}.
\begin{example}[$N=3$, $s=1$]
We consider the system, where 
the system should tolerate up to $s=1$ straggler. For the simplicity, we assume that the computation speeds of the three VMs are  $s[1] \ll s[2]=s[3]$. 
Note that the identity of the straggler is not known before the transmission. 
Hence, each dateset which is  assigned to only one VM in $[3]$ will not be considered into the transmission. In other words, we only consider the datesets in $\Ac_{\Vc}$ where $\Vc \subseteq [3]$ and $|\Vc|>1$.
  Denote $W_{\Vc}$ as the sum of the messages in $\Ac_{\Vc}$.
  For each $\Vc \subseteq [3]$ and $|\Vc|=2$, to tolerate $1$ straggler, $W_{\Vc}$ should be completely computed by the VMs in $\Vc$. 
  For $\Vc \subseteq [3]$ and $|\Vc|=3$, i.e., $\Vc=\{1,2,3\}$, we should determine the computation assignment by solving an optimization problem (see the optimization problem in~\eqref{eq: uncoded opt s} of Appendix~\ref{sec:general straggler}). Different from the orignal optimization problem in~\eqref{eq: uncoded opt 2},   the main difference  of the optimization problem in~\eqref{eq: uncoded opt s}   is that each dataset should be computed totally $s+1$ times, instead of $1$. In this example, since $s[1] \ll s[2]=s[3]$, 
we let $W_{\{1,2,3\}}$ completely computed by the workers in $\{2,3\}$.

  Then based on the computation assignment, we apply the coding technique in~\cite{heterogeneous_optimal}, to let VMs $V_1, V_2, V_3$ transmit 
  \begin{align*}
     & T_1 = \frac{1}{2}W_{\{1,2\}} + W_{\{1,3\}},\\
     &T_2 = \frac{1}{2}W_{\{1,2\}} + W_{\{2,3\}} + W_{\{1,2,3\}},\\
     &W_{\{1,3\}} - W_{\{2,3\}} - W_{\{1,2,3\}}, 
  \end{align*}
respectively. It can be seen that,
from any $2$ coded messages of $T_1, T_2, T_3$, the server can recover $W_{\{1,2\}} + W_{\{1,3\}} + W_{\{2,3\}} + W_{\{1,2,3\}}$. 
\hfill $\square$

\end{example}

\appendices

\clearpage
\bibliographystyle{IEEEtran}
\bibliography{references_d2d}

\clearpage

\section{Proof of Lemma~\ref{lem:existence of $c^*$}}
\label{appendix: exsitence}
There always exists $n^*$  which satisfies the Condition~\eqref{cond:2}, in the worst case, $n^* = 1.$ Condition~\eqref{cond:1} demonstrates that through the Algorithm~\ref{algorithm:lemma1}, the last $N - n^*$ VMs still have less computation time than $V_{n^*}.$ In another word, the computation load of the first $n^{*}$ VMs does not need to be rearranged since the computation time of the last $N - n^*$ VMs is strictly smaller than the over-all computation time $c^{\star}.$ In the worst case, $n^* = N.$ 

There are two situations in final computation load assignment,  $t_i > \frac{|\Lc_{i+1}/\Lc_{i-1}|}{s[i]+s[j]} > t_{i+1},$ and $t_i = \frac{|\Lc_{i+1}/\Lc_{i-1}|}{s[i]+s[j]} = t_{i+1}, for  i \in [n^*+1, N].$ In the second situation, $t_{n^*} > \frac{|\Lc_{i+1}/\Lc_{i-1}|}{s[i]+s[j]} = t_{i+1}$ from Condition~\eqref{cond:1}. In the first situation, $t_{n^*} > \frac{|\Lc_{i+1}/\Lc_{i-1}|}{s[i]+s[j]} = t_{i+1}$ from Condition~\eqref{cond:1}. 

\section{Converse Proof for Theorem~\ref{thm: explicit decentralized}}
\label{sec:converse proof}

The converse bound for  Theorem~\ref{thm: explicit decentralized} is proved as follows. Consider any integer $n^* \in [N]$. Since the datasets in $\Lc_{n^*}$ can only be computed by the VMs in $[n^*]$, thus by the cut-set strategy, the overall computation time $c$
 should be no less than the overall computation time if the system only needs to compute the sum of the messages in $\Lc_{n^*}$, which is no less than $\frac{|\Lc_{\Vc}|}{s[1]+\cdots+s[n^*]}$.\footnote{\label{foot:no less}This is becasue for any non-negative numbers $a_1,\ldots,a_n$ and $b_1,\ldots,b_n$, we have 
\begin{align*}
   & \frac{a_1+\cdots+a_n}{b_1+\cdots+b_n}=\frac{a_1}{b_1} \frac{b_1}{b_1+\cdots+b_n} +\cdots+\frac{a_n}{b_n} \frac{b_n}{b_1+\cdots+b_n} \\&
   \leq 
\max(\frac{a_1}{b_1},\ldots,\frac{a_n}{b_n}).
\end{align*}
 }
 Thus 
 \begin{align}
     c \geq \frac{|\Lc_{\Vc}|/K}{s[1]+\cdots+s[n^*]}, \ \forall n^* \in [N].
     \label{eq:one cut}
 \end{align}
Consider another cut of VMs,  $[n^*+1:n]$, where $n\in [n^*+1:N]$.     Since the datasets in $\Lc_{n}\setminus \Lc_{n^*}$ can only be computed by the VMs in $[n^*+1:n]$, thus by the cut-set strategy, the overall computation time $c$
 should be no less than the overall computation time if the system only needs to compute the sum of the messages in $\Lc_{n}\setminus \Lc_{n^*}$, which is no less than $\frac{|\Lc_{n}\setminus \Lc_{n^*}|/K}{s[n^*+1]+\cdots+s[N]}$. Thus 
  \begin{align}
     c \geq \frac{|\Lc_{n}\setminus \Lc_{n^*}|/K}{s[n^*+1]+\cdots+s[N]}, \ \forall n\in [n^*+1:N].
     \label{eq:sec cut}
 \end{align}
By~\eqref{eq:one cut}, ~\eqref{eq:sec cut}, and Lemma~\ref{lem:existence of $c^*$}, we can directly obtain the converse bound for Theorem~\ref{thm: explicit decentralized}.

\section{General Algorithm}
\label{appendix: alg}
\paragraph{Algorithm~\ref{algorithm:decentralized}}
The algorithm of the computation load assignment consists of  $N$ iterations  to calculate $\Sc[n,\Vc], \forall n \in [N], \forall \Vc \subset [N].$ We denote $\Sc[n] = \bigcup_{\Vc \subset [N]} \Sc[n,\Vc].$

\begin{algorithm} 
  \caption{Computation Load Assignment for Heterogeneous Computation Speed}
  \label{algorithm:decentralized}
  \begin{algorithmic}[1]
  \item[ {\bf Input}: $\sv$, $N$, $\alpha$ ] 
   \hspace*{4cm} 
  \For {$n \in [N]$}
  \State \quad $t_n \leftarrow \frac{|\Lc_{n} \setminus \Lc_{n-1}|}{s[n]} \quad $ set the computation time $t_n$.
  \State {\bf if $t_n \le t_{n-1}$} 
  \State \quad $\Sc[n] \leftarrow \cup_{\Vc \subset [n]: n \in \Vc} \Ac_{n,\Vc} \quad$ set the computation load of $V_n$. 
  \State {\bf else $t_n > t_{n-1}$} 
  \State \quad $n' \leftarrow n \quad$ preliminary setting of $n'$
  \State \quad $V_{n'-1}, V_{n'-2}, \ldots, V_{x}$ find the VMs that $t_{n'-1} = \ldots = t_{x}.$
  \State \quad  $\star$  update $\{\Sc[n], \ldots, \Sc[x]\}$ through Algorithm~\ref{algorithm:lemma1}.
  \State \quad {\bf if $  t_{x-1}\le \frac{|\Lc_{n} \setminus \Lc_{x-1}|}{\sum^{n}_{i=x} s[i]}$}
  \State \quad \quad  $n \leftarrow n+1$ 
  \State \quad {\bf else $\frac{|\Lc_{n} \setminus \Lc_{x-1}|}{\sum^{n}_{i=x} s[i]} < t_{x-1}$}
  \State \quad \quad  $n' \leftarrow x $ \quad {\bf reset $n'$, and return to $\star$.} 
  \State \quad {\bf end if}
  \State  {\bf end if}
  \EndFor
  \item[ {\bf Output}: $\Sc{[n]},$ $ n \in {[N]}$ ]
  \end{algorithmic}
\end{algorithm}


{\bf With no rearrangement (Line 8, 9)}
For each iteration $n \in N$, in Line $7$ we take $\Lc_{n+1} \setminus \Lc_{n}$ and $V_{n+1}$ into consideration, and we first let $V_{n+1}$ compute the $\Lc_{n+1} \setminus \Lc_{n}$ by itself, when $t_{n+1} \le t_{n},$ we update $\Sc[i]= \Sc[i], \forall i \in [n], \Sc[n+1] = \Lc_{n+1} \setminus \Lc_{n},$ and $t_{n+1} = \frac{|\Lc_{n+1} \setminus \Lc_{n}|}{s[n+1]}.$

{\bf With rearrangement (Line 11 $\sim$ 19)}
When $t_{n+1} > t_{n},$ we try to update $t_{n+1} = t_{n}.$
Firstly, we identify the VMs $\left\{V_{n-1}, V_{n-2}, \ldots, V_{x} \right\}$ that have the same computation time as $V_{n-1}.$ Next, we update $\{\Sc[n], \Sc[n-1], \ldots, \Sc[x]\}$ using Algorithm~\ref{algorithm:lemma1}. If $x == 1,$ we stop the rearrangement phase, if $x \neq 1,$ we repeat the process until $  t_{x-1} \le \frac{|\Lc_{n} \setminus \Lc_{x-1}|}{\sum^{n}_{i=x} s[i]}$.
  
This algorithm iteratively updates the computation load assignment by considering the difference in load between consecutive time steps. It adjusts $t_{n+1}$ based on the comparison with $t_{n}$, and if necessary, applies the load balancing scheme in Algorithm~\ref{algorithm:lemma1} to equalize the computation times. After $N$ iterations, the algorithm converges to the final load assignment that minimizes the overall computation time $c^{\star}$ with the algorithm complexity $\Oc(N)$.

\paragraph{Algorithm~\ref{algorithm:lemma1}}
Algorithm~\ref{algorithm:lemma1} provides the detailed rearrangement for the computation load assignment corresponding to {\bf Line 13} in Algorithm~\ref{algorithm:decentralized}. Where $d,l,m$ are indices to the VMs machines. $t_m = \frac{|\Lc_{d+l+m}- \Lc_{d+l+1}|}{s[d+l+1]+\ldots +s[d+l+m]}$ and $t_l =\frac{|\Lc_{d+l}- \Lc_{d+1}|}{s[d+1]+\ldots +s[d+l]}.$

\begin{algorithm} 
  \caption{ Computation Load Rearrangement }
  \label{algorithm:lemma1}
  \begin{algorithmic}[1]
  \item[ {\bf Input}: $\boldsymbol{s}$, $d$, $l$, $m$, $\Sc{[n]}$ $,n \in {[d+1 : d+l+m]} $, $t_m > t_l$, $\alpha$ ] 
   \hspace*{4cm} 
  \State  $\delta \leftarrow (\frac{|L_{l+m}|}{\sum_{i=1}^{l+m}s[i]} - t_{l})\sum_{i=1}^{l}s[i]$ calculate the size of rearranged computation load.
  \For {$n \in [d+1:d+l]$}
  \State  $\Sc[n,\Vc \cup \Qc] \leftarrow \frac{\delta |\Ac_{\Qc}|\mu[n,\Vc]}{|L_l||L_m||\Ac_{\Vc \cup \Qc}|}\Ac_{\Vc \cup \Qc} $, $ \forall \Vc \subset [l], \Qc \subset [m]$
  \EndFor
  \For {$n \in [d+l+1:d+l+m]$}
   \State  $\Sc[n,\Qc\bigcup \Vc] \leftarrow \Sc[n,\Qc \cup \Vc] \setminus \frac{\delta|\Ac_{\Vc}| \mu[n,\Qc]}{|L_l||L_m||\Ac_{\Qc \cup \Vc}|}\Ac_{\Qc \cup \Vc} $, $\forall \Qc \subset[m],  \Vc \subset [l]$
  \EndFor
  \item[ {\bf Output}:  $\Sc{[n]},$ $ n \in {[N]}$ ]
  \end{algorithmic}
\end{algorithm}

Firstly, we calculate the average computation time $t_{l+m} = \frac{|{\Lc}_{d+l+m} \setminus \Lc_{d+1}|}{\sum\limits_{n \in [d+1:d+l+m]} s[n]}$ and then we calculate the size of the exchanged computation load $\delta = (t_{l+m} - t_{l})\sum\limits_{n \in [d+1:d+l]} s[n].$ 

The key technology is to divide the rearranged load of $\{V_{n}, n \in [d+1:d+l]\}$ and $\{V_{n}, n \in [d+l+1:d+l+m]\}$ into two kinds of divisions:
\begin{subequations}
    \begin{align}
    \Dc &= \{\Ac'_{ \Vc \cup \Qc}, \forall \Vc \subset [d+1:d+l], \forall \Qc \subset [d+l+1:d+l+m]\} \\
    &= \{\Ac'_{ \Vc'},  \forall \Vc \subset [d+1:d+l]\} \\  \label{eq: division1}
    &= \{\Ac'_{ \Qc'}, \forall \Qc \subset [d+l+1:d+l+m]\},  
\end{align}
\end{subequations}
where $\Ac'_{\Vc \cup \Qc}$ denotes the rearranged part of $\Ac_{\Uc \cup \Vc \cup \Qc},$ $\Uc =\{ \emptyset, \Pc ,\forall \Pc \subset [d] \},$ $\Ac'_{\Vc'} = \{\Ac'_{\Vc \cup \Qc}, \forall \Qc \subset [d+l+1:d+l+m]\},$ $\Ac'_{\Qc'} = \{\Ac'_{\Vc \cup \Qc}, \forall \Vc \subset [d+1:d+l]\}.$ 

From the division~\eqref{eq: division1}, we find that $\Ac'_{\Vc}$ is one-to-one corresponding to $\Ac_{\Vc}$, as well as $\Ac'_{\Qc'}$ and $\Ac_{\Qc}$,
so we set $|\Ac'_{\Vc'}| = \frac{|\Ac_{\Vc}|}{|\Lc_{l}|}\delta,$ and $|\Ac'_{\Qc'}| = \frac{|\Ac_{\Qc}|}{|\Lc_{m}|}\delta.$ We design the rearrangement part $\Ac'_{n, \Vc \cup \Qc}$  corresponding to $\Sc[n,\Vc]$ and one-by-one. and so do to $\Ac'_{n, \Qc \cup \Vc}.$ To be specifically, 
$|\Ac'_{n, \Vc \cup \Qc}| = \frac{\mu[n,\Vc]|\Ac_{\Qc}|}{|\Lc_{l}||\Lc_{m}|}\delta, \forall n \in [d+1:d+l]$ and $|\Ac'_{n, \Qc \cup \Vc}| = \frac{\mu[n,\Qc]|\Ac_{\Vc}|}{|\Lc_{l}||\Lc_{m}|}\delta, \forall n \in [d+l+1:d+l+m].$ Finally, we update $t_{l+m}$ and $\Sc{[n]}, \forall n \in [d+1:d+l]$ in {\bf Line 5 \& 8} in Algorithm~\ref{algorithm:lemma1}.

\section{Experiment Results}
\begin{figure}
\centering
\includegraphics[width=0.75\columnwidth]{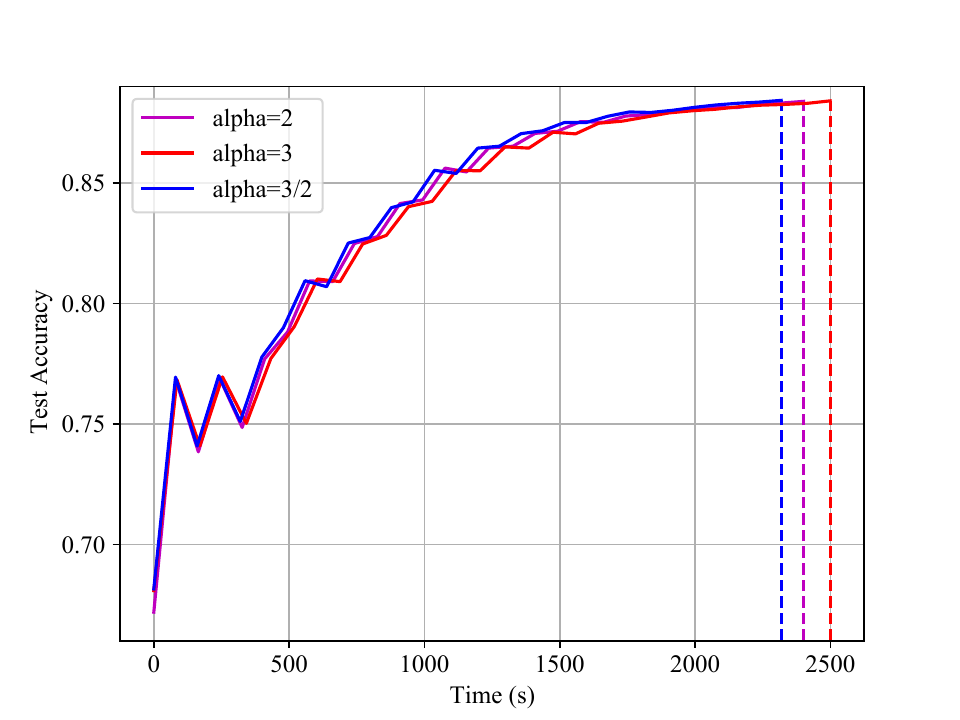}
\caption{ Results using different $\alpha$ with  DUSEC on Tencent cloud platform. 
}
\label{fig: Alpha}
\end{figure}
We also conduct a comparison among different $\alpha$ in DUSEC in terms of the accuracy and computation time, the experimental results indicate that when there is a large number of available VMs, the storage constraints have a minimal impact on the convergence accuracy. 

\section{General Description on the Proposed DUSEC Scheme against Stragglers}
\label{sec:general straggler}
To tolerate up to $s$ stragglers, computation redundancy is required. We   choose a redundancy parameter $m\geq 1$, and only focus on the datasets in $\Ac_{\Vc}$, where $|\Vc| \ge s+m,$ where each dateset should be computed by any $s+m$ VMs in $\Vc$. It will be clarifed later that $m$ will also play an important role on the number of transmisions by each VM, as shown in~\cite{heterogeneous_optimal}. 
The combinatorial convex optimization problem can be written as,
\begin{subequations} \label{eq: uncoded opt s}
\begin{align}
\underset{{\boldsymbol{M}} }{\text{minimize}} & \quad c\left(\boldsymbol{M}\right) = \max_{n \in  N}  \frac{\sum_{\Vc \subseteq [N]: n\in \Vc}\mu[n,\Vc]}{s[n]} \label{eq: uncoded s} \\
\text{subject to:}   
& \quad \sum_{n \in \Vc} \mu[n,\Vc] = (s+m)|\Ac_{\Vc}|,  \forall \Vc \subset [N], |\Vc|\geq s+m \\
& \quad 0 \leq  \mu[n,\Vc] \leq |\Ac_{\Vc}|, \forall n \in N.
\end{align}
\end{subequations}
When we solve the above problem, we can use the special polynomial coding in~\cite{heterogeneous_optimal} to let each VM transmit a coded message with $\frac{1}{m}$ length of the original message. 
In the following, we will propose a novel low-complexity algorithm to achieve the optimal solution for this  
problem. Interestingly, the {\em filling algorithm} introduced in the 
CSEC framework with heterogeneous computation speed \cite{woolsey2021cec} or the heterogeneous storage-constrained private information retrieval problem \cite{woolsey2021pir} can be applied here with small modifications to obtain the proposed optimal solution.

\begin{algorithm} 
  \caption{Adaptive Straggler Tolerant Decentralized Uncoded Storage Elastic Computing}
  \label{algorithm:2}
  \begin{algorithmic}[1]
  \item[ {\bf Input}: $\hat{\boldsymbol{s}}$, $\gamma$, $S$, $T$, $\beta^{t}$ ] 
   \hspace*{4cm} 
   \State $\boldsymbol{\nu}\leftarrow \hat{\boldsymbol{s}}$: same for all worker VMs
  \For {$t \in [T]$}
   \State {\bf At Master Machine}:
  \State \quad $\hat{\boldsymbol{s}}\leftarrow \gamma \boldsymbol{\nu} + (1-\gamma)\hat{\boldsymbol{s}}$  { (update estimate of speed vector)}.
  \State \quad $\mathcal{N}_t \leftarrow$ list of available machines 
  \State \quad 
  $\{F_g, \boldsymbol{\Mc}_g, \boldsymbol{\Pc}_g: \forall g \in [G]\} \leftarrow$ Results of computation assignment algorithm for $\Xm_g$ with straggler tolerance of $S$ for available machines $\mathcal{N}_t$ with speeds of $\hat{\boldsymbol{s}}$
  
  \State \quad Send $\beta_t$ and 
  $\{F_g, \boldsymbol{\Mc}_g, \boldsymbol{\Pc}_g: \forall g \in [G]\}$ to worker VMs
   \State {\bf At Worker VMs}:
   \State \quad $n\leftarrow$ index of worker VM
   \State \quad $\mu[n] \leftarrow$ total computation load of worker VM $n$
  \State \quad $c^{\star}_1\leftarrow$ current time
  \State \quad Perform assigned computations based on  $\{F_g, \boldsymbol{\Mc}_g, \boldsymbol{\Pc}_g: \forall g \in [G]\}$ 
  \State \quad $c^{\star}_2\leftarrow$ current time
  \State \quad $\nu[n] \leftarrow \mu[n]/(c^{\star}_2-c^{\star}_1)$ {(calculate speed based on current time step)} 
  \State \quad Send computations and $\nu[n]$ to Master Machine
  \State {\bf At Master Machine}:
  {after receiving results from at most $N - S$ workers}.
  \State \quad $\beta_{t+1}\leftarrow$ Combine worker results
  \EndFor
  \item[ {\bf Output}: $\beta^{t}$ ]
  \end{algorithmic}
\end{algorithm}


\begin{algorithm}
  \caption{Computation Load Assignment for $\boldsymbol{M}_t$ for 
  Heterogeneous Computation Speed}
  \label{algorithm:1}
  \begin{algorithmic}[1]
  \item[ {\bf Input}: $\boldsymbol{\mu}_g^\star$, 
  $q$, $\boldsymbol{\Zc}$ and $\Nc_g = \{1,\cdots,N_g\}$. ]
  \item $\boldsymbol{m} \leftarrow \boldsymbol{\mu}_g^\star$
  \item $f \leftarrow 0$
  \While {$\boldsymbol{m}$ contains a non-zero element}
    \State $f \leftarrow f+1$
    \State $L' \leftarrow \sum_{i=1}^{N_g}m[i]$
    \State $N'\leftarrow$ number of non-zero elements in $\boldsymbol{m}$
    \State $\boldsymbol{\ell} \leftarrow$ indices that sort the non-zero elements of $\boldsymbol{m}$ from smallest to largest\footnotemark[3]
    \State $\mathcal{P}_{g,f} \leftarrow\{\ell [1], \ell [N'-L+2] , \ldots , \ell [N'] \}$
    \If {$N' \geq L+1$}
    \State $\alpha_{g,f} \leftarrow  \min \left(\frac{L'}{L} - m[\ell[N' - L + 1]], m[\ell[1]]\right)$\footnotemark[4] 
    \Else
    \State $\alpha_{g,f} \leftarrow  m[\ell[1]]$
    \EndIf
    \For {$n \in \mathcal{P}_{g,f}$}
    \State $m[n] \leftarrow m[n] - \alpha_{g,f}$
    \EndFor
  \EndWhile
  \item $F \leftarrow f$
  \State Partition rows $[\frac{q}{G}]$ of $\Xm_g$ into $F$ disjoint row sets 
  $\mathcal{M}_{g,1}, \ldots , \mathcal{M}_{g,F}$ of size $\frac{\alpha_1 q}{G},\ldots,\frac{\alpha_{F}q}{G}$ rows, respectively
  \item[ {\bf Output}: $F$, 
  $\{ \mathcal{M}_{g,1}, \ldots , \mathcal{M}_{g,F}\}$ and 
  $\{ \mathcal{P}_{g,1}, \ldots , \mathcal{P}_{g,F}\}$ ]
  \end{algorithmic}
\end{algorithm}

The coding part is based on the polynomial codes~\cite{heterogeneous_optimal}, which can be written in a Matrix form,
\begin{align}
    AB &= \left[MDS\right] \times \left[\begin{array}{c}
         (Deamnd)  \\
         (Variable matrix) 
    \end{array}\right] =  C \\
    &= \left[     \begin{array}{ccccc}
         \star & \star & 0 & \ldots & 0  \\
         0 & \star & \star & \ldots & 0  \\ 
         \vdots & \vdots & \vdots & \ddots & \vdots \\
         0 & 0 & 0 & \ldots & \star
    \end{array}    \right],
\end{align}
where $A$ is a MDS matrix constituted from polynomial codes~\cite{heterogeneous_optimal} and Matrix $B$ contains the demand $(1,1,\ldots,1)$ in the first line. Matrix $C$ shows the the gradient sets that each VM can transmit. From $m$ coded messages of any $N - s$ VMs, it is possible to  recover the task.  

Consider a system with parameters $(K = 16000, N = 4, \alpha = 2, \sv = [1,2,5,5], s=1, m=2),$ and the optimal computation load are $\mu[1] = \cup_{\Vc \subset [4]: 1\in \Vc,|\Vc| \neq 4} \Ac_{\Vc},$$\mu[2] = \cup_{\Vc \subset [4]: 2\in \Vc} \Ac_{\Vc},$$\mu[3] = \cup_{\Vc \subset [4]: 3\in \Vc} \Ac_{\Vc},$$\mu[4] = \cup_{\Vc \subset [4]: 4\in \Vc} \Ac_{\Vc}$. We divide the gradients into non-overlap and equal length $2$ sub-vectors, the notation is simplified as $W_{\Vc} = \{W_{\Vc}(1) ,W_{\Vc}(2) \},$ where $W_r$ presents the gradients in $\Ac_{\Vc}.$ We only consider the gradients in $\Ac_{\Vc}, |\Vc| \ge 3$ in the coding phase for simplicity.

As illustrate in Fig.~\ref{fig: coded_computing}, $V_1$ transmits coded messages with half length 
$T_1 = - 3W_{\{1,2,3\}}^{(1)} - 2W_{\{1,2,4\}}^{(1)} -W_{\{1,3,4\}}^{(1)} - W_{\{1,2,3\}}^{(2)} - 2W_{\{1,2,4\}}^{(2)} +W_{\{1,3,4\}}^{(2)}. $ As well as $T_2 = - 2W_{\{1,2,3\}}^{(1)} -W_{\{1,2,4\}}^{(1)} +W_{\{2,3,4\}}^{(1)} - 2W_{\{1,2,3\}}^{(2)} - 3W_{\{1,2,4\}}^{(2)} - W_{\{2,3,4\}}^{(2)} + W_{\{1,2,3,4\}}^{(1)} - W_{\{1,2,3,4\}}^{(2)},$ $T_3 = - W_{\{1,2,3\}}^{(1)} +W_{\{1,3,4\}}^{(1)} +2W_{\{2,3,4\}}^{(1)} + W_{\{1,2,3\}}^{(2)} + 3W_{\{1,2,4\}}^{(2)} + 2W_{\{2,3,4\}}^{(2)} + 2W_{\{1,2,3,4\}}^{(1)} + 2W_{\{1,2,3,4\}}^{(2)},$ and $T_4 =  W_{\{1,2,4\}}^{(1)} +2W_{\{1,3,4\}}^{(1)} + 3W_{\{2,3,4\}}^{(1)} - W_{\{1,2,4\}}^{(2)} + 2W_{\{1,3,4\}}^{(2)} + W_{\{2,3,4\}}^{(2)} + \frac{1}{2}W_{\{1,2,3,4\}}^{(1)} - \frac{1}{2}W_{\{1,2,3,4\}}^{(2)}.$
In matrix form,
\begin{align}
    \left[ T_1; T_2; T_3; T_4; T_5\right] = \left[ \begin{array}{cccccc}
         -3 & -2 & -1 & 0 & 0  & 1 \\
         -2 & -1 & 0 & 1 & 1  & -2 \\
         -1 & 0 & 1 & 2 & 2  & 1 \\
         0 & 1 & 2 & 3 & 0.5  & 0 \\
         -1 & 2 & -1 & -0.5 & 0 & 0 \\
    \end{array} \right] \times \left[ \begin{array}{c}
         W_{\{1,2,3\}}^{(1)} \\
         W_{\{1,2,4\}}^{(1)} \\
         \vdots \\
         W_{\{1,2,3,4\}}^{(1)} \\
         W_{\{1,2,3\}}^{(2)} \\
         \vdots \\
         W_{\{1,2,3,4\}}^{(2)}  
    \end{array}\right].
\end{align}
From any $3$ coded messages $T_n, n \in [4],$ the master can recover the task through coding. For example, when the master receives $T_1,T_2,T_3,$ it can code $\frac{1}{4}T_3 - \frac{3}{4}T_1 + \frac{1}{2}T_2 = \sum_{\Vc \subset [4]: |\Vc| \ge 3}W_{\Vc}(1),$ and $\frac{1}{4}T_3 + \frac{1}{4}T_1 - \frac{1}{2}T_2 = \sum_{\Vc\subset [4]: |\Vc| \ge 3}W_{\Vc}(2).$

\end{document}